\documentclass[12pt]{iopart}
\usepackage[dvips]{graphicx}
\usepackage{iopams} 
\usepackage{bm}

\def\rev#1{}

\newtheorem{definition}{Definition}

\newtheorem{lemma}{Lemma}
\newtheorem{theorem}{Theorem}

\newtheorem{aproblem}{Problem}

\newtheorem{remark}{Remark}

\def\cl#1{{\cal#1}}
\def\tr{{\rm Tr}}
\def\ket#1{|#1\rangle}
\def\bra#1{\langle#1|}

\def\nashi#1{}

\begin{document}
\title[LOCC-detection of a maximally entangled state]
{A study of LOCC-detection of a maximally entangled state
using hypothesis testing}
\author{Masahito Hayashi${}^1$
	Keiji Matsumoto${}^{2,1}$
	and
	Yoshiyuki Tsuda${}^3$\footnote{
The current address of the third author 
is Institute of Statistical Mathematics, 4-6-7 Minami-Azabu, Minato-ku, Tokyo 106-8569, Japan.}
}
\address{${}^1$ ERATO Quantum Computation and Information Project,
	JST, 5-28-3-201 Hongo, Bunkyo-ku, Tokyo 113-0033, Japan}
\ead{masahito@qci.jst.go.jp}
\address{${}^2$ National Institute of Informatics,
	2-1-2 Hitotsubashi, Chiyoda-ku, Tokyo 101-8430, Japan}
\ead{keiji@nii.ac.jp}
\address{${}^3$ COE, Chuo University,
	1-13-27 Kasuga, Bunkyo-ku, Tokyo 112-8551, Japan}
\ead{y-tsuda@ism.ac.jp}
\begin{abstract}
We study how well answer to the question
``Is the given quantum state 
equal to a certain maximally entangled state?''
using LOCC,
in the context of hypothesis testing.
Under several locality and invariance conditions,
optimal tests will be derived for several special cases
by using basic theory of group representations.
Some optimal tests are realized by 
performing quantum teleportation and checking 
whether the state is teleported.
We will also give a finite process for realizing some optimal tests.
The performance of the tests will be numerically compared.
\end{abstract}
\submitted{\JPA}
\pacs{03.65.Wj,03.65.Ud,02.20.-a}
%02.20.-a 	Group theory 
%(for algebraic methods in quantum mechanics, see 03.65.Fd; 
%for symmetries in elementary particle physics, see 11.30.-j)
%03.65.Ta Foundations of quantum mechanics; measurement theory 
%(for optical tests of quantum theory, see 42.50.Xa)
%03.65.Ud Entanglement and quantum nonlocality 
%(e.g. EPR paradox, Bell's inequalities, GHZ states, etc.) 
%(for entanglement production in quantum information, see 03.67.Mn; 
%for entanglement in Bose-Einstein condensates, see 03.75.Gg)
%03.65.Wj State reconstruction, quantum tomography
%42.50.-p 	Quantum optics
\maketitle

\section{Introduction}

Entanglement plays an important role in quantum information
\cite{bbcjpw,bennett-wiesher,ekert,nielsen-chuang}.
An experimental system
makes use of a certain maximally entangled state $\ket{\phi^0_{AB}}$
for realization of quantum information processing.
However,
a state generated as a maximally entangled state
is not necessarily a true maximally entangled state
because the entanglement is easily corrupted by interaction
with the environment. Hence, it is important
to consider how well answer to the question
`Is the state equal to $\ket{\phi^0_{AB}}$?'
using quantum measurement with two outcomes
$(T_0,\,T_1)$ corresponding to (Yes, No).

For the practical use,
it is natural to restrict
our measurements to Local Operation and Classical Communications (LOCC)
because LOCC are easily implemented.
Since the result of the LOCC measurement is probabilistic
and the error of incorrect answers is inevitable,
it is important to consider an optimization problem of the
measurement.
As a framework of this argument,
hypothesis testing
is appropriate \cite{hayashi-text}.
We consider two hypotheses
\begin{center}
	$H_0$: the state is $\ket{\phi^0_{AB}}$
	versus
	$H_1$: the state is not $\ket{\phi^0_{AB}}$
	.
\end{center}
Since $H_0$ is an accumulation point of $H_1$,
the probability of the correct answer
`$H_0$ is true' when $H_0$ is really true
is almost equal to the probability of the incorrect answer
`$H_0$ is true' when the state is close to $\ket{\phi^0_{AB}}$
but different.
In the hypothesis testing,
considering the two errors:
\smallskip\\
(i) To answer `$H_1$ is true' though $H_0$ is really true,
\\
(ii) To answer `$H_0$ is true' though $H_1$ is really true,
\smallskip\\
we minimize the probability of (ii) with the probability of
(i) kept small.
See Section 2 for details.

There are similar studies
based on entanglement witness;
A physical observable which gives minus outputs
for a set of entangled inputs
\cite{horodecki3,terhal}.
The concept of entanglement witness is
widely adopted,
and there are many extensive arguments,
especially, by making use of group symmetry.
See, for example, theoretical works
\cite{lewenstein,guhne,d'ariano}
and experiments
\cite{barbierietal}.
However,
in their arguments,
analysis of statistical error is not sufficient.
%Moreover,
%the method for $n$-sample case has not been
%developed.
Hence it is worth considering this problem in the style of
statistical hypothesis testing
\cite{lehmann}.
Though
there have been studies of
quantum hypothesis testing
\cite{helstrom,holevo,
hiai-petz,ogawa-nagaoka,hayashi-hypo,
asympt,hayashi-text},
there have not been enough arguments
for testing entanglement.

In this article,
we give an approach to
the hypothesis testing whether the state is $\ket{\phi^0_{AB}}$
using LOCC measurement
between two parties and independent samples.
We will derive optimal LOCC tests
under some group invariance.
The first case we consider is
testing one sample of a pair of $d$-dimensional systems using
LOCC between two parties.
As a physical meaning,
the optimal test is equivalent to
optimal teleportation using a given state partially entangled,
and
the error probability is the same as
the fidelity of the input and the output of the teleportation.
It is also found that the test
is equivalent to the extreme points of LOCC measurement
described by Virmani and Plenio \cite{virmani-plenio}
and the entanglement witness given in \cite{d'ariano}.
Next,
the result is generalized for the $n$-sample case.
We derive an optimal test which is invariant by
$SU(d^n)$,
and its asymptotic behavior ($n\to\infty$).
In the asymptotic sense,
the optimal LOCC test has the same performance as
the optimal test without LOCC restriction.
Next,
we present the main result of this article:
For $d=n=2$,
the optimal test using LOCC between
parties and independent samples,
with $SU(2)$-invariance and some additional conditions
or requirements.
Since these tests are characterized by 
invariant measure,
it contains continuous operations,
However, in order to implement it, they need their construction with
finite basis.
Then, we show how to construct the optimal measurement
with finite basis, for experimental realization.
Finally,
we consider an optimal test using measurement
non-local between samples
with $SU(2)\times SU(2)$-invariance.
This test is equivalent to the entanglement swapping.

This article is organized as follows.
In Section 2,
a general formulation of hypothesis testing is introduced.
In Section 3,
we state problems treated in this article.
In Section 4,
we consider a problem to test entanglement based on
a single sample pair,
and we derive an optimal test $T^u$.
Moreover,
we consider a case where there are
$n$-independent pairs of samples to test entanglement.
As a direct consequence of the previous section,
we derive an optimal LOCC test $T^U$.
It is also shown that this test has the same performance as
the optimal test without LOCC restriction in an asymptotic sense.
%In Section 5,
%we consider a case where we can exploit
%two independent pairs of samples.
In Section 5,
an optimal test $T^V$ is derived under an LOCC condition
between AB-parties and between samples.
In Section 6,
an optimal test $T^W$ is also derived
under another condition which is less restrictive
as for locality.
%In Section 7,
%we discretize $T^u$ and $T^V$ using representation theory
%of a finite group.
In Section 7,
we discretize the test derived in Section 5
using representation of a finite groups.
In Section 8,
we compare the performance of these tests
for $n=2$.

\section{Hypothesis testing}\label{sec:hypo}

The main subject of this article is to test
whether a given state is
\begin{center}
	$H_0$: the maximally entangled state $\ket{\phi^0_{AB}}$
	or
	$H_1$: any other state
\end{center}
using LOCC.
To setup the hypothesis testing formally,
we first consider hypotheses $H_0$ and $H_1$
generally consisting of many elements.
The hypothesis testing is an optimization problem
with respect to error probability
of a measurement with two outcomes
corresponding to the two hypotheses.
As described later, there are two error probabilities,
and one of them will be minimized
with the other one kept in a given small level.

Let ${\cl H}$ be a finite-dimensional Hilbert space
which describes a physical system of interest.
We denote 
the set of linear operators (matrices) on ${\cl H}$
(of density matrices on ${\cl H}$)
by $\cl L(\cl H)$ and $\cl S(\cl H)$,
respectively.
In hypothesis testing,
we assume two hypotheses 
the {\it null hypothesis} $H_0$ 
and the {\it alternative hypothesis} $H_1$,
and choose 
two non-empty subsets 
${\cl S}_0$ and ${\cl S}_1$
of ${\cl S}({\cl H})$
such that ${\cl S}_0\cap{\cl S}_1=\emptyset$,
which correspond to our hypotheses.
Suppose that the given state $\rho(\in{\cl S}({\cl H}))$ of
the system is unknown and that
$\rho\in{\cl S}_0$ or $\rho\in{\cl S}_1$.
We test
\begin{equation}
	\label{eq:hypo}
	H_0:\rho\in{\cl S}_0\mbox{ versus }H_1:\rho\in{\cl S}_1
\end{equation}
by a measurement with two outcomes $T=(T_0,T_1)$:
If the outcome $T_i$ is obtained,
then we support the hypothesis $H_i$.
However, the purpose of hypothesis testing is 
rejecting the null hypothesis $H_0$ 
and accepting $H_1$ with a given confidence level.
Hence, we make decision only when the outcome $T_1$ is observed,
and we reserve our decision when the outcome $T_0$ is observed.
For simplicity, the test, or the measurement,
$T$ is often described by $T_0$.
In the hypothesis testing, 
there are two kinds of errors:
Type 1 error is an event such that
$H_1$ is accepted though $H_0$ is true.
Type 2 error is an event such that
$H_0$ is accepted though $H_1$ is true.
Hence
the type 1 error probability $\alpha(T,\rho)$
and
the type 2 error probability $\beta(T,\rho)$
are given by
\[
	\alpha(T,\rho)=\tr(\rho T_1) \ (\rho\in{\cl S}_0),\
	\beta (T,\rho)=\tr(\rho T_0) \ (\rho\in{\cl S}_1).
\]
A test $T$ is said to be {\it level-$\alpha$}
when $\alpha(T,\rho)\le\alpha$ for any $\rho\in{\cl S}_0$
because $\alpha$ expresses the confidence level of our decision.
A quantity $1-\beta(T,\rho)$ is called {\it power}.
In our main problem,
we will consider level-zero tests only.

The main problem of hypothesis testing is
to maximize the power,
or equivalently, to minimize the type 2 error probability,
of the test $T$ of level-$\alpha$.
A test $T$ of level-$\alpha$ is said to be
{\it Most Powerful (MP) level-$\alpha$}
at $\rho\in{\cl S}_1$
if $\beta(T,\rho)\le\beta(T',\rho)$
for any level-$\alpha$ test $T'$.
A test $T$ of level-$\alpha$ is said to be
{\it Uniformly Most Powerful (UMP) level-$\alpha$}
if $T$ is MP level-$\alpha$ for any $\rho\in{\cl S}_1$.
The UMP test is regarded as the best test.
However, except for some examples,
there is no UMP test because the uniformness is
too strict.

In mathematical statistics, 
it is too difficult to solve problems when
both ${\cl S}_0$ and ${\cl S}_1$ have plural elements,
except for some special cases,
for example, the classical bioequivalence problem \cite{bhm}.
Hence, it is natural to consider the case where
\[
	{\cl S}_0:=\{\bra{\phi^0_{AB}}\rho\ket{\phi^0_{AB}}\le c\}
	,\
	{\cl S}_1:=\{\bra{\phi^0_{AB}}\rho\ket{\phi^0_{AB}} > c\},	
\]
or
\[
	{\cl S}_0:=\{\rho\neq \ket{\phi^0_{AB}}\bra{\phi^0_{AB}}\}
	,\
	{\cl S}_1:=\{\rho= \ket{\phi^0_{AB}}\bra{\phi^0_{AB}}\}.
\]
However, it is also too difficult to treat the above case.
Hence, we consider the case
${\cl S}_0:=\{\ket{\phi^0_{AB}}\bra{\phi^0_{AB}}\}$
in this article.

If any $\rho\in{\cl S}_0\cup{\cl S}_1$
is invariant by an action of a group,
e.g., transposition of the order of independent samples,
we can without loss of generality
restrict attention to tests exhibiting
the same invariance,
because the error probabilities are invariant.
We may also require that $T_0$ should be invariant
by a group action leaving ${\cl S}_0$ invariant
to simplify the problem mathematically.
In experiments of entanglement,
only LOCC can be used,
so it is required that the test is realized by LOCC.

There is a trade-off between requirement and power of a test;
If there are two requirements $C_1$ and $C_2$ for a test
and if $C_1$ is weaker than $C_2$,
then the optimal test for $C_1$ is more powerful than
that for $C_2$.
If $C_1$ and $C_2$ are unitary-invariance conditions,
arguments for $C_2$ tends to be mathematical easier.
If they are locality conditions,
arguments for $C_2$ tends to be more difficult.
In the next section, we will introduce some different
conditions.

\section{Problems treated in this article}
\rev{13}

Suppose that
$n$-independent samples
are provided, that is, the state is given in the form
\begin{equation}
	\label{eq:state}
	\rho=
	\sigma^{\otimes n}=
	\underbrace{\sigma\otimes\cdots\otimes\sigma}_{n}
\end{equation}
for an unknown density $\sigma$ of a single sample.
We test the following hypothesis with level zero:
\begin{equation}
	\label{eq:our:hypo}
	H_0:\sigma  =  \ket{\phi^0_{AB}}\bra{\phi^0_{AB}}
	\mbox{ versus }
	H_1:\sigma \ne \ket{\phi^0_{AB}}\bra{\phi^0_{AB}}
	.
\end{equation}
Here,
\[
	\ket{\phi^0_{AB}}=
	\frac{1}{\sqrt d}\sum_{i=0}^{d-1}
	\ket{i}_{A}\otimes\ket{i}_{B}
\]
is a vector of a maximally entangled pair
on two $d$-dimensional parties A and B spanned by
$\ket0_A,\ket1_A,...,\ket{d-1}_A$
and
$\ket0_B,\ket1_B,...,\ket{d-1}_B$, respectively.
We refer to $\{\ket i_A\}$ and $\{\ket i_B\}$
as the {\it standard basis}.

Since the state is invariant
by transposing the order of independent samples,
we can without loss of generality impose that
the tests for each case should be invariant by the same transposition.
We additionally impose three types of basic conditions on tests,
that is,
level-zero, locality and unitary invariance.
Among various level-$\alpha$ conditions,
we adopt $\alpha=0$ because it is the most fundamental
and the optimal tests have
analytically simple forms.
We use only {\it AB-local} tests,
i.e., LOCC between A and B.
In some cases,
we also require that tests should be
{\it samplewise-local}, i.e., LOCC between independent samples.
Unitary invariance of the measurement is imposed
for the symmetry of $\sigma^{\otimes n}$
or $(\ket{\phi^0_{AB}}\bra{\phi^0_{AB}})^{\otimes n}$.

First, we will make an LOCC test for a product system
of two $d$-dimensional systems.
Then, this will be generalized to the case of
$n$-independent pairs of the $d\times d$ systems.
For the $n$-sample case,
a samplewise locality condition can be considered.
Without the samplewise locality,
we will derive optimal tests for any $d$ and $n$.
With the samplewise locality, however,
the problem is so difficult that
we will derive optimal tests only for $d=n=2$.

We list three sets of conditions under which we will find
best tests in Sections 4-6.
Unless otherwise mentioned, AB-locality is always imposed.

\begin{remark}\rm
One may think that
it is impossible to prepare the plural samples of 
the given unknown state $\sigma$ 
when the state is easily corrupted by interaction.
However, 
the density $\sigma$ represents the ensemble of 
states generated by a specific state generator.
Hence, as long as each sample is generated by this generator,
it can be regarded as the state $\sigma$.
\end{remark}

\subsection{U-invariance for $n$-samples}
As an action of $SU(d^n)$,
{\it U-action} $U_{A B}$ is defined as
\begin{equation}
	\label{eq:u-inv}
%	U_{A B}(g)^\dagger
%	T_0\,
%	U_{A B}(g)
%	\quad
	U_{A B}(g)=U_A(g)\otimes\overline U_B(g)
	\mbox{ for }
	g\in SU(d^n)
\end{equation}
where
$U_A$ and $U_B$ are the natural representations
of $SU(d^n)$ on the $d^n$-dimensional subsystems A and B,
respectively, and $\overline X$ is the contragradient of $X$
with respect to the standard basis,
i.e., $(\overline X)_{i,j}= X_{j,i}$.
The state $\ket{\phi^0_{AB}}\bra{\phi^0_{AB}}$ is U-invariant in the sense
$U_{A B}(g)\ket{\phi^0_{AB}}\bra{\phi^0_{AB}}U_{A B}^\dag(g)
=\ket{\phi^0_{AB}}\bra{\phi^0_{AB}}$.
A test $T=(T_0,T_1)$ is said to be U-invariant
if $T_0=U_{A B}^\dag(g) T_0U_{A B}(g)$.
Under the AB-locality condition,
a UMP U-invariant test $T^U$
will be derived.
Moreover,
it will be shown that, asymptotically, $T^U$ has the
same performance as a test which is UMP
without the AB-locality or the U-invariance
(Section 4).

\subsection{Samplewise-locality and V-invariance
for two samples}

Let $d=n=2$.
We require
samplewise-locality, that is, in this case,
a test $T$ is realized by LOCC
between the first and the second samples.
The {\it V-action} of $SU(2)$ is defined as
\begin{equation}
	\label{eq:v-inv}
	V_{A_1 B_1 A_2 B_2}:=
	U_{A_1}\otimes \overline{U_{B_1}}
	\otimes U_{A_2}\otimes \overline{U_{B_2}}
	.
\end{equation}
In the same sense as the U-invariance,
$(\ket{\phi^0_{AB}}\bra{\phi^0_{AB}})^{\otimes2}$ is V-invariant.
Moreover, V leaves the set ${\cl S}_1$ invariant
while U and W (defined below) do not.
A test is said to be V-invariant if
$V_{A_1 B_1 A_2 B_2}^\dag T_0V_{A_1 B_1 A_2 B_2}$
is invariant.
The V-invariance is not so strict as the U-invariance
that its mathematical analysis is difficult.
Hence we also consider {\it AB-invariance}.
This is invariance by {\it AB-transpositions} generated by
\begin{equation}
	\label{eq:ab-trans}
	\ket{i}_{A_1}\ket{j}_{B_1}\ket{k}_{A_2}\ket{l}_{B_2}
	\mapsto
	(-1)^{i+j+k+l}
	\ket{1-j}_{A_1}\ket{1-i}_{B_1}\ket{1-l}_{A_2}\ket{1-k}_{B_2}
	.
\end{equation}
A UMP V-invariant test $T^V$ will be derived
under the samplewise-locality,
the AB-invariance,
and {\it termwise AB-covariance} defined in
Definition \ref{def:termwise}.
Moreover,
it will be shown that
in a subset of density operators,
$T^V$ is UMP without the termwise AB-covariance
(Section 5).

\subsection{W-invariance for two samples}
Let $d=n=2$ again.
The {\it W-action} of the direct product
$SU(2)\times SU(2)$ is defined as
\begin{equation}
	\label{eq:w-inv}
	W_{A_1 B_1 A_2 B_2}(g,h):=
	U_{A_1}(g)\otimes\overline{U_{B_1}}(g)\otimes
	U_{A_2}(h)\otimes\overline{U_{B_2}}(h)
\end{equation}
for $g,h\in SU(d)$.
$\ket{\phi^0_{AB}}\bra{\phi^0_{AB}}$ is again W-invariant,
and a test $T=(T_0,T_1)$
is said to be W-invariant if
$W_{A_1 B_1 A_2 B_2}^\dag T_0 W_{A_1 B_1 A_2 B_2}$
is invariant.
The W-invariance is weaker than the U-invariance
but is stronger than the V-invariance.
In a subset of density operators,
a UMP W-invariant test $T^W$
is obtained
(Section 6).

The U-invariance is the most strict condition and
the V-invariance is the weakest
as the $SU(2)$ action for $d=n=2$.
As for the locality conditions,
the samplewise-locality in addition to the AB-locality
treated in the V-invariance case is the most strict.
As is shown in  Section \ref{sec:dis} with a graph,
the power of $T^W$ is the highest,
that of $T_U$ is the second and
that of $T^V$ is the lowest in a neighborhood of $H_0$.
Hence
it is recommended to use $T^W$ rather than $T^U$
when one can use non-local measurement between the
two independent samples.
However, asymptotically, $T^U$ is optimal.
See Section \ref{sec:asympt}.

\section{U-invariance}

In this section, as the first step,
we consider the case of $n=1$.
As the next step,
we generalize the result to arbitrary $n$.

\subsection{One-sample case}
Let $n=1$.
Virmani and Plenio \cite{virmani-plenio}
have derived extreme points of AB-local measurements
using Positive Partial Transpose (PPT).
We will derive the same measurement $T^u=\{T_0^u,T_1^u\}$
as a UMP U-invariant test
using property of separable measurement.

\begin{theorem}
For $n=1$,
a UMP AB-local and U-invariant  test $T_0^u$ of level-zero
is given as follows:
\begin{equation}
	\label{eq:th1:1}
	T_0^u=
	\ket{\phi^0_{AB}}\bra{\phi^0_{AB}}
	+
	\frac{1}{d+1}
	(I-\ket{\phi^0_{AB}}\bra{\phi^0_{AB}})
	.
\end{equation}\label{th:1}
The type 2 error probability is
\begin{equation}
	\label{eq:th1:2}
	\beta(T_0^u,\rho)=\beta(T^u,\sigma)=
	\frac{d\theta+1}{d+1},
\end{equation}
where $\theta=\bra{\phi^0_{AB}}\sigma\ket{\phi^0_{AB}}$.
\end{theorem}

The formula (\ref{eq:th1:2}) shows that
the power of the test goes to zero
as the state goes to $\ket{\phi^0_{AB}}$.
Hence, it is difficult to reject $H_0$ even if $H_1$ is true.
The case when $\sigma$ is in a neighborhood of $H_0$
will be highlighted in (\ref{9-6-3}) and (\ref{9-6-4}) in the next subsection.
%the forcoming paper \cite{newhayashi}.
Other optimal tests derived in the later sections
have the same property.

\begin{remark}
\rm
The protocol for the test $T^u$ is implemented using the teleportation.
Suppose that Alice has a state $\ket\psi$ in another system $A'$.
She measures her total system $A\otimes A'$ by the Bell basis
and then she lets Bob know the result.
The teleportation is completed when Bob rotates the system
according to the Alice's information.
The imperfectness causes some error in the teleportation,
and the fidelity 
$|\langle \psi\ket{\psi'}|^2$ of the teleported state
$\ket{\psi'}$ is evaluated by
the measurement $\{\ket\psi\bra\psi,I-\ket\psi\bra\psi\}$,
This process is equivalent to the test $T^u$ with $A'$ ignored,
and the fidelity is the same as $\beta(T_0^u,\rho)$.
\end{remark}

\begin{remark}
\rm
Virmani and Plenio \cite{virmani-plenio}
has proved that $T^u$ is an extreme point
of AB-local measurements under invariance conditions.
Their work is related to our problem since
an optimal test is always an extreme point
though the converse is not always true.
In the case $n=1$,
they found that there are two extreme points.
As a test, however,
it is obvious that the measurement other than $T^u$
is not optimum as a test for the hypothesis.
Hence we can also conclude that $T^u$ is optimum
based on their approach.

D'Ariano \etal \cite{d'ariano} have also considered
the same measurement as $T^u$,
as an entanglement witness.
However,
it is different from the hypothesis testing because
the optimization of the error probability
was not considered.
\end{remark}
\medskip
\noindent
{\bf Proof of Theorem \ref{th:1}} \ \
First,
we show that 
$T_0^u$ can be
written as a classical mixture
of AB-local projective measurements, i.e.,
\begin{eqnarray}
\fl	T_0^u=
	\int_{g\in SU(d)}
	(U_A(g)\otimes \overline U_B(g))^\dagger
	\Big(
	\sum_{i=1}^d
	\ket{i}_{A} \ket{i}_{B}
	\bra{i}_{A} \bra{i}_{B}
	\Big)
	(U_A(g)\otimes \overline U_B(g))
	\mu(d g),\label{9-6}
\end{eqnarray}
where $\mu(\cdot)$ is the Haar measure on $SU(d)$.
(Its full measure is $1$.)
From the invariance, we can easily see that
the LHS has the form
$a \,\ket{\phi^0_{AB}}\bra{\phi^0_{AB}}
+b (I-\ket{\phi^0_{AB}}\bra{\phi^0_{AB}})$.
Since 
\begin{eqnarray}
\Tr (\hbox{LHS of }(\ref{9-6}) )= 
d \hbox{ and }
\bra{\phi^0_{AB}} (\hbox{LHS of }(\ref{9-6}) )
\ket{\phi^0_{AB}}= 1,
\end{eqnarray}
we obtain (\ref{9-6}).

Then, the test
\begin{eqnarray*}
& (U_A(g)\otimes \overline U_B(g))^\dagger
	\Big(
	\sum_{i=1}^d
	\ket{i}_{A} \ket{i}_{B}
	\bra{i}_{A} \bra{i}_{B}
	\Big)
	(U_A(g)\otimes \overline U_B(g))\\
=&
	\sum_{i=1}^d
	U_A(g)^\dagger \ket{i}_{A} \overline U_B(g)^\dagger \ket{i}_{B}
	\bra{i}_{A} U_A(g)
	\bra{i}_{B}\overline U_B(g)
\end{eqnarray*}
can be realized by 
the local measurements based on the bases
$\{U_A(g)^\dagger \ket{i}_{A} \}_{i=1}^d$
and $\{\overline U_B(g)^\dagger \ket{i}_{B}\}_{i=1}^d$.
Hence, the test $T^u$ is realized by measuring $T=\{T_0,T_1\}$ 
by randomly choosing $g$ subject to the Haar measure.

Next, we prove its optimality.
A U-invariant test $T_0$ is written in the following form:
\[
	T_0=
	a \,\ket{\phi^0_{AB}}\bra{\phi^0_{AB}}
	+b (I-\ket{\phi^0_{AB}}\bra{\phi^0_{AB}})
\]
where $0\le a\le1$ and $0\le b\le1$.
Since
$\bra{\phi^0_{AB}}T_0\ket{\phi^0_{AB}}=a$,
$T_0$ is level-zero if and only if $a=1$.
Hence, it is sufficient to show that
any LOCC level-zero U-invariant test $T$ satisfies $b\ge(d+1)^{-1}$.
Further, since $a=1$, the condition $b\ge(d+1)^{-1}$
is equivalent with the condition 
\begin{eqnarray}
\Tr T_0 \ge d.\label{9-6-2}
\end{eqnarray}

Now, we will show (\ref{9-6-2}). 
Since an LOCC measurement is separable,
$T_0$ should be separable between A and B,
that is,
\[
	T_0=
	\sum_i c_i
	M_{A,i}\otimes M_{B,i}
\]
where $0\le c_i\le1$
and
where $M_{A,i}$ and $M_{B,i}$ are rank-one projections
on $A$ and $B$, respectively.
Since
$\tr(T_0)=a+b(d^2-1)=1+b(d^2-1)=\sum_i c_i$,
our problem is to minimize $\sum_i c_i$.
Let $F_i=\tr(M_{A,i} M_{B,i}^T)$,
where $X^T$ is the transpose of $X$ with respect to
the standard basis.
Then,
\begin{eqnarray*}
\fl	1=\bra{\phi^0_{AB}}T_0\ket{\phi^0_{AB}}
	=\sum_i c_i
	\bra{\phi^0_{AB}}
		M_{A,i}\otimes M_{B,i}
	\ket{\phi^0_{AB}}
	=\frac{\sum_i c_i \tr(M_{A,i} M_{B,i}^T)}{d}
	=\frac{\sum_i c_i F_i}{d}
	.
\end{eqnarray*}
Since $0\le F_i\le 1$, we have
\begin{equation}
	\label{eq:th1:trace}
	\tr(T_0)=\sum_i c_i \ge \sum_i c_i F_i =d.
\end{equation}
\hfill$\Box$

%\begin{remark}
%In (\ref{eq:th1:1}),
%$T_0^u$ is constructed as
%a classical mixture of continuously
%many local measurements.
%It means that, in experiments,
%we should deal with
%continuously many bases to measure the system,
%so it costs much  in experiments.
%To reduce the cost,
%we would like to realize
%$T_0^V$ as a mixture of finite elements
%if possible.
%In fact, it is possible since the test
%\end{remark}

The unconditionally UMP level-zero test $T_0^g$ is
$T_0^g=\ket{\phi^0_{AB}}\bra{\phi^0_{AB}}$,
and
its type 2 error is
$\beta(T_0^g,\sigma)=\theta$.
The AB-locality is reflected in
the difference $(1-\theta)/(d+1)$ of the type 2 errors
of $T_0^u$ and $T_0^g$.

\begin{remark}\rm
In order to prove the optimality,
we focused on the trace of $T_0$.
This trace method is very powerful for treating 
the separable POVM element
detecting a given entangled state with probability one.
This method was invented in this research for the first time,
and was applied to other papers \cite{owari,damian}.
\end{remark}

\subsection{\boldmath$n$-sample case}

Theorem 1 is generalized to the case of arbitrary $n$
as follows.
\begin{theorem}
For any $n\ge1$,
a UMP AB-local and U-invariant test of level-zero is
\[
	T_0^U=
	(\ket{\phi^0_{AB}}\bra{\phi^0_{AB}})^{\otimes n}
	+
	\frac{1}{d^n+1}(I-(\ket{\phi^0_{AB}}
	\bra{\phi^0_{AB}})^{\otimes n})
	.
\]
The type 2 error probability is
\[
	\beta(T_0^U,\sigma^{\otimes n})=
	\frac{d^n\theta^n+1}{d^n+1}
\]
where $\theta=\bra{\phi^0_{AB}}\sigma\ket{\phi^0_{AB}}$.
\end{theorem}
{\bf Proof} \ \
The proof of Theorem 1 is directly applied
by replacing
the space $A$ in Theorem 1 with $A_1\otimes\cdots\otimes A_n$,
$B$ with $B_1\otimes\cdots\otimes B_n$,
the dimension $d$ with $d^n$
and the group $SU(d)$ with $SU(d^n)$.
\hfill$\Box$

\subsection{Asymptotic property}\label{sec:asympt}
For comparison,
let us consider other tests:
\[
	T_0^{u,n}=(T_0^u)^{\otimes n}
	\quad \mbox{ and }\quad
	T_0^G=(T_0^g)^{\otimes n}
	=(\ket{\phi^0_{AB}}\bra{\phi^0_{AB}})^{\otimes n}
	.
\]
Note that
they are both level-zero since $T_0^u$ and $T_0^g$ are
level-zero.
We also note that
$T_0^G$ is UMP level-zero without any condition.
The type 2 error probabilities are
\[
	\beta(T_0^{u,n},\sigma^{\otimes n})=
	\Big(\frac{d\theta+1}{d+1}\Big)^n,\
	\beta(T_0^G,\sigma^{\otimes n})=
	\theta^n.
\]
Hence we have
\[
	\beta(T_0^G,\sigma^{\otimes n})
	<
	\beta(T_0^U,\sigma^{\otimes n})
	<
	\beta(T_0^{u,n},\sigma^{\otimes n})
	\quad
	(n\ge2)
	.
\]
On the other hand,
the asymptotic behavior of
$\beta(T_0^U,\sigma^{\otimes n})$ is
\begin{eqnarray}
	\lim_{n\to\infty}
	\frac{\beta(T_0^U,\sigma^{\otimes n})}{\theta^n}=1
	& \mbox{ if } \theta\ge1/d,\label{9-6-3} \\
	\lim_{n\to\infty}
	\frac{\beta(T_0^U,\sigma^{\otimes n})}{1/d^n}=1
	& \mbox{ if } \theta<1/d.\label{9-6-4}
\end{eqnarray}
It means that,
if $\theta=\bra{\phi^0_{AB}}\sigma\ket{\phi^0_{AB}}\ge1/d$
then
$T_0^U$ and $T_0^G$
have the same asymptotic performance
not only for the exponent but also for the coefficient
of the type 2 error probabilities.
In this sense,
the restriction of AB-locality and U-invariance
does not reduce
the performance of the UMP level-zero test $T_0^G$.

\section{Samplewise-locality,
V-invariance for \boldmath$n=d=2$}

We consider the case $n=d=2$.
First, we derive a UMP test $T^V$
under the conditions of
samplewise-locality, V-invariance, AB-invariance,
and the {\it termwise AB-covariance}
(defined in Definition \ref{def:termwise}).
We then prove that this test is
also UMP
without the termwise AB-covariance
for a subset $\cl S'$ of density operators.

Before defining termwise covariance,
we note that,
if $T_0$ is AB-local and samplewise-local
then $T_0$ is AB-separable and samplewise-separable, that is,
\[
	T_0=
	\sum_i
	p_i
	M_{A_1,i} \otimes M_{B_1,i}
	\otimes
	M_{A_2,i} \otimes M_{B_2,i}
\]
where $M_{X}$ is a rank-one projection on the system $X$.

\begin{definition}\label{def:termwise}
\rm
The test $T_0$ is said to be
{\it termwise AB-covariant} if
\[
	\tr(M_{A_1,i}\overline{M_{B_1,i}^T})=1
	\mbox{ and }
	\tr(M_{A_2,i}\overline{M_{B_2,i}^T})=1
\]
holds.
\end{definition}
The meaning of the termwise AB-covariance will be clarified
by Hayashi \cite{newhayashi}.

\rev{4}
Define $\ket{\phi^1_{AB}}$, $\ket{\phi^2_{AB}}$ and
$\ket{\phi^3_{AB}}$ as follows:
\begin{eqnarray*}
	&&
	\ket{\phi^1_{AB}}:=
	\frac{\sqrt{-1}}{\sqrt{2}}
	\big(
	 \ket{0}_{A}\otimes\ket{1}_{B}
	+\ket{1}_{A}\otimes\ket{0}_{B}
	\big),
	\\
	&&
	\ket{\phi^2_{AB}}:=
	\frac{1}{\sqrt{2}}
	\big(
	-\ket{0}_{A}\otimes\ket{1}_{B}
	+\ket{1}_{A}\otimes\ket{0}_{B}
	\big),
	\\
	&&
	\ket{\phi^3_{AB}}:=
	\frac{\sqrt{-1}}{\sqrt{2}}
	\big(
	 \ket{0}_{A}\otimes\ket{0}_{B}
	-\ket{1}_{A}\otimes\ket{1}_{B}
	\big).
\end{eqnarray*}
In this section,
we frequently use the matrix expression
$x_{ij}=\bra{\phi^i_{AB}}\sigma\ket{\phi^j_{AB}}$
for the sake of notational convenience.

\begin{remark}\label{rem:so3}
\rm
There is a two-to-one group homomorphism
of $SU(2)$ onto $SO(3)$ as the three-dimensional subrepresentation of
$U_A\otimes\overline{U_B}$.
%leaving $\ket{\phi^0_{A B}}$ invariant and
It irreducibly acts on
${\rm span}\{
	\ket{\phi^1_{A B}},
	\ket{\phi^2_{A B}},
	\ket{\phi^3_{A B}}
\}$.
Now, we regard the tensor product space 
$
({\rm span}\{
        \ket{\phi^0_{A B}}, 
	\ket{\phi^1_{A B}},
	\ket{\phi^2_{A B}},
	\ket{\phi^3_{A B}}
\})^{\otimes 2}$ as the space $M$ of $4\times4$ matrices
spanned by the basis $e^{i j}:=\ket{\phi^i_{AB}}_1\ket{\phi^j_{AB}}_2$.
$SU(2)$ acts on $M$ by $V_{A_1 B_1 A_2 B_2}$
as follows:
\begin{equation}
	\left[
	\begin{array}{c|c}
	1 & 0 \\\hline
	0 & S
	\end{array}
	\right]
	\left[
	\begin{array}{c|ccc}
	e^{00} & e^{01} & e^{02} & e^{03} \\\hline
	e^{10} & e^{11} & e^{12} & e^{13} \\
	e^{20} & e^{21} & e^{22} & e^{23} \\
	e^{30} & e^{31} & e^{32} & e^{33} \\
	\end{array}
	\right]
	\left[
	\begin{array}{c|c}
	1 & 0 \\\hline
	0 &  S^T
	\end{array}
	\right]
	\label{eq:so3}
\end{equation}
for $S\in SO(3)$.
Let $K_i^\pm$, $L_i^\pm$ be $i$-dimensional subspaces of $M$
defined as follows:
\smallskip\\
$K_6^+$: The space of all $3\times3$ symmetric matrices
spanned by $e^{i j}$ $(1\le i,j\le3)$,
\\
$K_1^+$: The one-dimensional subspace of $K_6^+$
spanned by the $3\times3$ identity matrix,
\\
$K_3^+$: The three-dimensional subspace of $K_6^+$
spanned by
$e^{i j}+e^{j i}$ $(i\ne j)$,
\\
$K_2^+$: The two-dimensional space spanned by
\[
	e^{11}+\omega e^{22}+\omega^2 e^{33}
	\mbox{ and }
	e^{11}+\omega^2 e^{22}+\omega e^{33}
\]
where $\omega$ is a solution to $\omega^2+\omega+1=0$,
\\
$K_5^+:= K_6^+ - K_1^+=K_3^+ + K_2^+$,
\\
$K_3^-$: The space of all $3\times3$ alternating matrices
spanned by $e^{i j}$ $(1\le i,j\le3)$,
\\
$M_{10}^+$: The ten-dimensional space of
all $4\times4$ symmetric matrices,
\\
$M_6^-$: The six-dimensional space of
all $4\times4$ alternating matrices,
\\
$L_1^+$: The one-dimensional space spanned by
$e^{00}=\ket{\phi^0_{AB}}^{\otimes2}$,
\\
$L_3^+:=M_{10}^+ - K_6^+ - L_1^+$,
\\
$L_3^-:=M_6^- - K_3^-$.
\smallskip\\
The V-action
$V=U_{A_1}\otimes \overline{U_{B_1}}\otimes U_{A_2}
\otimes\overline{U_{B_2}}$
is equivalent to
$U_{A_1}\otimes U_{B_1}\otimes U_{A_2} \otimes U_{B_2}$
as group representation.
By the V-action
(or, equivalently,
by the SO(3) action of the form (\ref{eq:so3})),
$M$ is decomposed into subspaces of
irreducible representations as
\begin{equation}
	M=K_5^+
	\underbrace{\oplus K_3^- \oplus L_3^+ \oplus L_3^-}
	_{{\rm equivalent}}
	\underbrace{\oplus K_1^+ \oplus L_1^+}
	_{{\rm equivalent}}
	.
	\label{eq:so3:deco}
\end{equation}
%The irreducibility of this decomposition is proved by
%the representation theory of
%symmetric groups and general linear groups.
See \cite{fulton,goodman} for details.
The decompositions into the three spaces
$K_3^-$ and $L_3^\pm$
and into the two spaces
$K_1^+$ and $L_1^+$ in (\ref{eq:so3:deco}) are not unique
because they have the equivalent
representations of
three-dimension and one-dimension, respectively.

The AB-transposition
simultaneously maps
$\ket{\phi^0_{AB}}_i$ to $-\ket{\phi^0_{AB}}_i$
for $i=1,2$,
while it leaves other $\ket{\phi^i_{AB}}_i$ invariant.
Hence it acts on $M$ as
\begin{equation}
	M\ni X\mapsto
\left(
\begin{array}{cc}
 -1 & 0 \\ 0 & I_3 
\end{array}\right)
	X
\left(	
\begin{array}{cc}
 -1 & 0 \\ 0 & I_3 
\end{array}
\right)
	\label{eq:ab-inv:mat}
\end{equation}
where $I_3$ is the three-dimensional identity matrix.
Hence it makes $-1$-multiplication on $L_3^\pm$
while $K_3^-$ is left invariant.
Transposition of the order of the independent samples
corresponds to the matrix transposition of $M$.
Hence it makes $-1$-multiplication on $K_3^-$ and $L_3^-$
while $L_3^+$ is left invariant.
Therefore,
by the V-action with the two types of transposition,
$M$ is decomposed as
\[
	M=
	K_5^+\oplus
	\underbrace{K_3^-\oplus L_3^+\oplus L_3^- \oplus}
	_{{\rm not\ equivalent}}
	\underbrace{K_1^+\oplus L_1^+}
	_{{\rm equivalent}}
	.
\]

By the W-action,
$M$ is decomposed as
\[
	M=L_1^+
	\oplus
	L_3'
	\oplus
	L_3''
	\oplus
	K_9
\]
where
$L_3'$ and $L_3''$ are the three-dimensional spaces spanned by
$x_{i,0}$ and $x_{0,j}$, respectively,
and
$K_9$ is the nine-dimensional space spanned by
$x_{i,j}$ $(1\le i,j\le 3)$.
Though $L_3'$ and $L_3''$ has the same dimension,
this decomposition is unique because
the first and the second element of $SU(2)\times SU(2)$
independently act on $L_3'$ and $L_3''$.
The transposition of the order of independent samples
corresponds to transposing $L_3'$ and $L_3''$.
Hence W-invariant test for $\sigma^{\otimes2}$
has the same weight on $L_3'$ and $L_3''$.
\end{remark}

\subsection{Termwise AB-covariance}

We use the symbols $K_i^\pm$ and $L_i^\pm$
not only as the spaces
but also as the projection operators.
Any operator $X$ invariant by the V-action, the AB-transposition,
and the transposition of the order of independent samples
is of the form
\[
	X=
	w_1 K_5^+ + w_2 K_3^-  + w_3 L_3^+ + w_4 L_3^-
	+ J
\]
where $0\le w_i\le1$
and $J$ is an operator on the two-dimensional space
$J_2:=K_1^+\oplus L_1^+$.
Each weight $w_i$ and the form of $N$ of the optimal test
for $\ket{\phi^0_{AB}}$ is obtained as follows.

\begin{theorem}\label{th:3}
A UMP
AB-local, samplewise-local, V-invariant, AB-invariant
and termwise AB-covariant test of level-zero is
given as
\begin{equation}
	\label{eq:th3:t0}
	T_0^V=
	 \frac1{10} K_5^+
	+\frac13 L_3^+
	+(\ket{\phi^0_{AB}}\bra{\phi^0_{AB}})^{\otimes2}
	+\frac16 K_3^-
	+\frac13 L_3^-
	.
\end{equation}
The type 2 error of $T_0^V$ is
\begin{equation}
	\label{eq:th3:beta}
	\beta(T_0^V,\sigma^{\otimes2})=
	v^T Z v-\frac2{15}
	({\rm Re}(x_{12})^2+{\rm Re}(x_{23})^2+{\rm Re}(x_{31})^2)
\end{equation}
where
\[
	v=\left(
	\begin{array}{c}
	x_{11}-1/2\\ x_{22}-1/2\\ x_{33}-1/2
	\end{array}\right),\
	Z=
	\frac1{15}
	\left(\begin{array}{ccc}
	6&7&7\\7&6&7\\7&7&6
	\end{array}\right)
	.
\]
\end{theorem}

\noindent
{\bf Proof of Theorem \ref{th:3}} \ \
First,
the all conditions of locality and invariance
are checked by calculating the weight for each projection of
\begin{equation}
	\label{eq:tv:twirl}
	T_0^V
	=
	\int_{g\in SU(2)}
	(V_{A_1 B_1 A_2 B_2}(g))^\dagger
	(\Pi_{00} + \Pi_{01} + \Pi_{10} + \Pi_{11} )
	(V_{A_1 B_1 A_2 B_2}(g))
	\mu(d g)
\end{equation}
where $\mu(\cdot)$ is the Haar measure on $SU(2)$ and where
$\Pi_{ij}$ ($i,j=0,1$) is the projection
on the one-dimensional subspace spanned by
\begin{equation}
	\label{eq:pi_ij}
	\ket i_{A_1}
	\otimes
	\ket i_{B_1}
	\otimes
	\frac{\ket0_{A_2} +(-1)^j \ket1_{A_2}}{\sqrt2}
	\otimes
	\frac{\ket0_{B_2} +(-1)^j \ket1_{B_2}}{\sqrt2}
\end{equation}
(see also Section \ref{sec:v-o} bellow).

%line feed
Next,
we show that the type 2 error of $T_0^V$ is minimized.
Any test satisfying all those conditions
is given in the form
\begin{eqnarray}
	T_0
	&=&
	\sum_i
	q_i
	\int_{g\in SU(2)}
	U_{A_1}(g)^\dagger\ket0_{A_1}\bra0_{A_1}U_{A_1}(g)
	\otimes
	U_{B_1}(g)^T\ket0_{B_1}\bra0_{B_1}\overline U_{B_1}(g)
	\nonumber
	\\
	&&
	\nonumber
%	\label{eq:t0:fid}
	\otimes
	U_{A_2}(g)^T
	\ket{\psi_{F_i}}_{A_2}\bra{\psi_{F_i}}_{A_2}
	\overline U_{A_2}(g)
	\otimes
	U_{B_2}(g)^T
	\ket{\psi_{F_i}}_{B_2}\bra{\psi_{F_i}}_{B_2}
	\overline U_{B_2}(g)
	\mu(d g)
\end{eqnarray}
where $q_i\ge0$ and
\[
	\ket{\psi_{F}}_X=
	\frac{\sqrt{F}\ket0_X + \sqrt{1-F}\ket1_X}{\sqrt2}
	\quad
	(0\le F\le 1)
	.
\]
%The form of (\ref{eq:t0:fid}) shows that $T_0$ is
%AB-invariant.
%Moreover,
%We can assume that $T_0$ is also {\it onetwo-invariant},
%that is, invariant by the permutation between
%$A_1\otimes B_1$ and $A_2\otimes B_2$,
%since the sample $\rho=\sigma^{\otimes2}$ is 12-invariant.
For the invariance conditions,
$T_0$ can be written as
\[
	T_0
	=
	  w_1 K_5^+ + w_2 L_3^+
	+ J
	+ w_3 K_3^- + w_4 L_3^-
	\quad
	(0\le w_i\le1)
	.
\]
\nashi
{
where $N$ is a positive operator on the two-dimensional
subspace $K_1^+ \oplus L_1^+$.
$N$ is of the form
\begin{equation}
	\label{eq:v:l=0}
	N=
	w_3 K_1^+ + L_1^+
\end{equation}
because of the level-zero condition
$\bra{\phi^0_{AB}}^{\otimes2}T_0\ket{\phi^0_{AB}}^{\otimes2}$.
}
To satisfy the level-zero condition,
the weight of $J$ for $L_1^+$ should be one.
To minimize the type 2 error,
the weight of $J$ for $K_1^+$ should be zero.
Hence $T_0$ should be
\[
	T_0
	=
	  w_1 K_5^+ + w_2 L_3^+
	+ L_1^+
	+ w_3 K_3^- + w_4 L_3^-
	.
\]
Define
\[
	m(X)=
	\bra0_{A_1}\bra0_{B_1}\bra{\psi_F}_{A_2}\bra{\psi_F}_{B_2}
	X
	\ket0_{A_1}\ket0_{B_1}\ket{\psi_F}_{A_2}\ket{\psi_F}_{B_2}
	.
\]
By direct calculation,
$m(X)$ is given as follows:
\begin{eqnarray*}
	&&
	m(K_5^+)
	=
	\frac{F^2-F+1}{6}
	,\
	m(L_3^+)
	=
	\frac{F}{2}
	,\
	m(K_1^+)
	=
	\frac{(2F-1)^2}{12}
	,\
	m(L_1^+)
	=
	\frac14,
	\\
	&&
	m(K_3^-)
	=
	\frac{F(1-F)}{2}
	,\
	m(L_3^-)
	=
	\frac{1-F}{2}
	.
	\\
	&&
%	m(N_0)
%	=
%	\sqrt{-1}\frac{2F-1}{4\sqrt3}
\end{eqnarray*}
\nashi
{
where
\[
	N_0
	=
	\frac
	{
	\ket{\phi^1_{AB}}_1\ket{\phi^1_{AB}}_2
	+
	\ket{\phi^2_{AB}}_1\ket{\phi^2_{AB}}_2
	+
	\ket{\phi^3_{AB}}_1\ket{\phi^3_{AB}}_2
	}
	{\sqrt3}
	\bra{\phi^0_{AB}}_1\bra{\phi^0_{AB}}_2
	.
\]
}
Moreover,
\begin{eqnarray}
	\tr(\sigma^{\otimes2}K_5^+)
	=&
	\frac1{3}
	(x_{11}+x_{22}+x_{33})^2
	+
	\frac1{6}
	\sum_{1\le i<j\le3} (x_{i i}+x_{j j})^2
	\nonumber
	\\
	&
	-
	\frac4{3}
	\sum_{1\le i<j\le3} ({\rm Im}x_{i j})^2
	+
	\frac1{3}
	\sum_{1\le i<j\le3} |x_{i j}|^2
	,
	\label{eq:tr:K_5^+}
	\\
	\tr(\sigma^{\otimes2}L_3^+)
	=&
	\sum_{i=1}^3
	\big( x_{00} x_{i i} +	|x_{0 i}|^2 \big)
	,
	\label{eq:tr:L_3^+}
	\\
	\tr(\sigma^{\otimes2}K_1^+)
	=&
	\frac13
	\sum_{1\le i,j\le3} x_{i j}^2
	,
	\label{eq:tr:K_1^+}
	\\
	\tr(\sigma^{\otimes2}L_1^+)
	=&
	x_{00}^2
	,
	\label{eq:tr:L_1^+}
	\\
	\tr(\sigma^{\otimes2}K_3^-)
	=&
	\sum_{1\le i<j\le3} x_{i i} x_{j j}
	-
	\sum_{1\le i<j\le3} |x_{i j}|^2
	,
	\label{eq:tr:K_3^-}
	\\
	\tr(\sigma^{\otimes2}L_3^-)
	=&
	\sum_{i=1}^3
	\big( x_{00} x_{i i} -	|x_{0 i}|^2 \big)
	.
	\label{eq:tr:L_3^-}
%	\\
%	\tr(\sigma^{\otimes2} N_0)
%	=&
%	\frac
%	{x_{01}^2 + x_{02}^2 + x_{03}^2}
%	{\sqrt3}
%	\label{eq:tr:N_0}
\end{eqnarray}
Hence,
the type 2 error probability
%with respect to
%$x_{ij}=\bra{\phi^i_{AB}}\sigma\ket{\phi^j_{AB}}$
is given by
\begin{equation}
	\label{eq:v:abc}
	\beta(T_0,\sigma^{\otimes2})
	=
	\sum_i q_i (a F_i^2 + b F_i + c)
%	=
%	\sum_i q_i a F_i^2 + 2 b + 4 c
\end{equation}
where
\begin{eqnarray*}
\fl	a
	&=&
	\frac
	{(x_{11}-x_{22})^2+(x_{22}-x_{33})^2+(x_{33}-x_{11})^2}
	{15}
	+
	\frac
	{2}{5}
	\big(
	{({\rm Re}x_{12})^2}
	+
	{({\rm Re}x_{23})^2}
	+
	{({\rm Re}x_{31})^2}
	\big)
	,
	\\
\fl	b
	&=&
	-a+
	\frac
	{({\rm Im}x_{01})^2+({\rm Im}x_{02})^2+({\rm Im}x_{03})^2}
	{6}
	,
	\\
\fl	c
	&=&
	-
	\frac
	{({\rm Im}x_{01})^2+({\rm Im}x_{02})^2+({\rm Im}x_{03})^2}
	{3}
	+
	\frac
	{({\rm Re}x_{12})^2+({\rm Re}x_{23})^2+({\rm Re}x_{31})^2}
	{15}
	+
	v_0^T Z_0 v_0
	,
\end{eqnarray*}
and where
\[
	v_0=\left(\begin{array}{c}
	x_{11}-1/2\cr x_{22}-1/2\cr x_{33}-1/2
	\end{array}\right)
	,\
	Z_0
	=
	\frac1{30}
	\left(\begin{array}{ccc}
	4 & 3 & 3 \cr 3 & 4 & 3 \cr 3 & 3 & 4
	\end{array}\right)
	.
\]
We minimize (\ref{eq:v:abc}) under
necessary conditions on $\sum_i q_i$ and $\sum_i q_i F_i$
as follows.
Since $T_0$ is level-zero, we have
\begin{equation}
	\label{eq:v:q_i}
	\frac14 \sum_i q_i
	=
	\bra{\phi^0_{AB}}^{\otimes2}
	T_0
	\ket{\phi^0_{AB}}^{\otimes2}
	=
	1
	.
\end{equation}
%From (\ref{eq:v:l=0}),
%$\sum_i q_i m(L_0) =0$,
We have
\begin{equation}
	\label{eq:v:q_i:2}
	\sum_i q_i F_i =2
	.
\end{equation}
Hence,
the type 2 error probability (\ref{eq:v:abc})
is minimized if $\sum_i q_i F_i^2$ is minimized
under (\ref{eq:v:q_i}) and (\ref{eq:v:q_i:2}).
From Jensen's inequality,
\[
	\sum_i q_i F_i^2
	=
	4
	\sum_i \frac{q_i}{4} F_i^2
	\ge
	4(\sum_i \frac{q_i}{4} F_i)^2
	=
	1.
\]
The equality holds if
$q_1=\cdots=q_4=1$ and $F_1=\cdots=F_4=1/2$
so that
the type 2 error probability is uniformly minimized
if $T_0=T_0^V$.
Hence we obtain (\ref{eq:th3:t0}) and (\ref{eq:th3:beta}).
\hfill$\Box$

\subsection{Optimality without termwise AB-covariance}

In this subsection,
we discuss the optimality of $T^V$ under another conditions,
removing the termwise AB-locality.
In this argument,
we use PPT instead of separability of measurement.
PPT is a class of tests which strictly includes the set of
separable/LOCC tests.
Hence, a test is best among LOCC
if it is LOCC and is best among PPT.
The set of PPT tests satisfies
some linear inequalities for weights
on projections $K^\pm_i$ and $L^\pm_i$.
So $T^V$ is optimal in PPT if
it uniformly minimizes error probability
under the condition of the linear inequalities\rev{6}.

We consider parameterized subsets of states as follows.

%$T_0^V$ is
%UMP AB-local samplewise-local V-invariant AB-invariant
%without the termwise AB-covariance condition
%in a set ${\cl S}(\theta_0)$ of density operators
%for $0\le\theta_0\le1$
%defined as follows.
%
\begin{definition}
\rm
Let $S(\vartheta)$ be a set of density operators $\sigma$
satisfying the following two conditions
for $x_{ij}=\bra{\phi^i_{AB}}\sigma\ket{\phi^j_{AB}}$:
\[
	\theta=x_{00}=\bra{\phi^0_{AB}}\sigma\ket{\phi^0_{AB}}\ge\vartheta,
\]
and
\begin{equation}
	\label{eq:ineqwithout}
	\frac12
	\sum_{1\le i<j\le3}
	(x_{ii}-x_{jj})^2
	+
	3
	\sum_{1\le i<j\le3}
	|x_{i j}|^2
	\ge
	4
	\sum_{1\le i<j\le3}
	({\rm Im}x_{i j})^2
	,
\end{equation}
or equivalently,
\begin{equation}
	\label{eq:s1>l3+}
	3\tr(\sigma^{\otimes2}K_1^+)
	\ge
	\tr(\sigma^{\otimes2}K_3^-)
	.
\end{equation}
\end{definition}
This condition
(\ref{eq:ineqwithout}) is satisfied if
\[
	\sigma=(1-p-q-r)\ket{\phi^0_{AB}}\bra{\phi^0_{AB}}
	+p\ket{\phi^1_{AB}}\bra{\phi^1_{AB}}
	+q\ket{\phi^2_{AB}}\bra{\phi^2_{AB}}
	+r\ket{\phi^3_{AB}}\bra{\phi^3_{AB}}
	.
\]
Indeed, it holds that
\[
	3\tr(\sigma^{\otimes2}K_1^+)
	-
	\tr(\sigma^{\otimes2}K_3^-)
	=
	\frac{(p-q)^2+(q-r)^2(r-p)^2}{2}
	\ge0
	.
\]

%Our hypothesis in this section is reformulated
%as follows:
%\begin{equation}
%	\label{eq:v-hypo}
%	H_0:\sigma  =  \ket{\phi^0_{AB}}\bra{\phi^0_{AB}}
%	\mbox{ versus }
%	H_1:\sigma \ne \ket{\phi^0_{AB}}\bra{\phi^0_{AB}}
%	\mbox{ and }
%	\sigma\in{\cl S}(\theta_0)
%	.
%\end{equation}

\begin{theorem}\label{th:opt}
There is $\theta_0 <1$ such that
$T_0^V$ is
UMP AB-local, samplewise-local, V-invariant,
weakly AB-invariant with level-zero
in $\cl S(\theta_0)$.
\end{theorem}
{\bf Proof} \ \
In this proof,
we deal with the alternative side
$T_1=I-T_0$ of the measurement
because it makes the calculation simple;
$T_1$ has the zero weight on $L_1^+$.
If $T_1$ satisfies
%AB-local, samplewise-local, V-invariant, AB-invariant
all the locality and invariance conditions
and if it is level-zero,
then $T_1$ is given by
\[
	T_1=
	w_1 K_5^+
	+
	w_2 L_3^+
	+
	w_3 K_1^+
	+
	w_4 K_3^+
	+
	w_5 L_3^-
	.
\]
%for the same reason in the proof of Theorem 3.
The power %, one minus the type 2 error probability,
of the test is given as
\begin{eqnarray*}
\fl	\tr(\sigma^{\otimes2}T_1)=
	w_1 \tr(\sigma^{\otimes2}K_5^+)
	+
	w_2 \tr(\sigma^{\otimes2}L_3^+)
	+
	w_3 \tr(\sigma^{\otimes2}K_1^+)
	+
	w_4 \tr(\sigma^{\otimes2}K_3^-)
	+
	w_5 \tr(\sigma^{\otimes2}L_3^-)
	.
\end{eqnarray*}
%Recall that
%$\sigma$ satisfies the condition (\ref{eq:v:<})
%for $\cl S(\theta_0)$.
Lemma \ref{lem:5w1+3w2} shows that,
if $1-\theta$ is small,
the power % (=one minus the type 2 error probability)
is maximized if
\begin{equation}
	\label{eq:5w1+3w2}
	5 w_1 + 3 w_2,\
	w_2,\
	w_3,\ 
	5 w_1 + 3 w_2 + w_3 + 3 w_4 + 3 w_5, \mbox{ and }
	w_5
\end{equation}
are simultaneously maximized.
From Lemmas \ref{lem:1}-\ref{lem:3} in Appendix,
$w_1,...,w_5$ should satisfy
\begin{eqnarray}
	&&
	\frac{10 w_1 + 6 w_2 - w_3}{12}\le1,
	\label{eq:v:1}
	\\
	&&
	\frac{w_3+2(w_4+w_5)}{4}\le1,
	\label{eq:v:2}
	\\
	&&
	w_2=w_5,
	\label{eq:v:3}
	\\
	&&
	\frac34(w_2+w_5)\le1.
	\label{eq:v:4}
\end{eqnarray}
Therefore,
\begin{eqnarray*}
	&&
	\max\{5 w_1 + 3 w_2
	\mid
	(\ref{eq:v:1}),0\le w_i \le1
	\}
	=
	\frac{13}{2},
	\\
	&&
	\max\{w_2
	\mid
	(\ref{eq:v:3}),(\ref{eq:v:4}),0\le w_i \le1
	\}
	=
	\frac{2}{3},
	\\
	&&
	\max\{w_3
	\mid
	(\ref{eq:v:1}),(\ref{eq:v:2}),0\le w_i \le1
	\}
	=
	1,
	\\
	&&
	\max\{ 5 w_1 + 3 w_2 + w_3 + 3 w_4 + 3 w_5
	\mid
	0\le w_i \le1\} = 12,
	\\
	&&
	\max\{w_5
	\mid
	(\ref{eq:v:3}),(\ref{eq:v:4}),0\le w_i \le1
	\}
	=
	\frac{2}{3},
\end{eqnarray*}
and we have (\ref{eq:th3:t0}) as a solution to the
linear maximization problem.
\hfill$\Box$

\section{W-invariance for \boldmath$n=d=2$}
Let $d=n=2$.
In this section,
we test the following hypothesis with level zero:
\begin{equation}
	\label{eq:w-hypo}
	H_0:\sigma=\ket{\phi^0_{AB}}\bra{\phi^0_{AB}}
	\mbox{ versus }
	H_1:
	1/4\le\bra{\phi^0_{AB}}\sigma\ket{\phi^0_{AB}}<1
	.
\end{equation}
In other words,
we consider the case where the set of possible states is
$\cl S'=\{\sigma\mid
\bra{\phi^0_{AB}}\sigma\ket{\phi^0_{AB}}\ge1/4\}$.
%The hypothesis is reformulated as
%\begin{equation}
%	\label{eq:w-hypo}
%	H_0:\sigma  =  \ket{\phi^0_{AB}}\bra{\phi^0_{AB}}
%	\mbox{ versus }
%	H_1:\sigma \ne \ket{\phi^0_{AB}}\bra{\phi^0_{AB}}
%	\mbox{ and }
%	\sigma\in{\cl S}
%	.
%\end{equation}

\begin{theorem}\label{th:w}
A UMP AB-local, and W-invariant
for (\ref{eq:w-hypo})
of level-zero is given as follows:
\begin{equation}
	\label{eq:th4:t0}
	T_0^W=
	\ket{\phi^0_{AB}}\bra{\phi^0_{AB}}^{\otimes2}+
	\frac13(I-\ket{\phi^0_{AB}}\bra{\phi^0_{AB}})^{\otimes2}
	.
\end{equation}
The type 2 error probability of $T_0^W$ is
\begin{equation}
	\label{eq:th4:beta}
	\beta(T_0^W,\sigma^{\otimes2})=
	\theta^2 + \frac{ (1-\theta)^2 }{3}
	.
\end{equation}
\end{theorem}

\begin{remark}
\rm
The test $T^W$ is implemented,
by using the entanglement swapping
from $A_1\otimes B_1$ and $A_2\otimes B_2$ to $B_1\otimes B_2$;
Measuring $A_1\otimes A_2$ in\rev{19} the Bell basis,
can create entanglement in $B_1\otimes B_2$.
The success rate, or the fidelity to the maximally entangled state,
of the swapping is
equivalent to the type 2 error probability  
$\beta(T_0^W,\sigma^{\otimes2})$.
\end{remark}
{\bf Proof of Theorem \ref{th:w}} \ \
$T_0$ is W-invariant (see Remark \ref{rem:so3}).
It is also AB-local because
\begin{eqnarray*}
	T_0^W
	=&
	\int_{g,h\in SU(2)}
	(W_{A_1 B_1 A_2 B_2}(g,h))^\dagger
	\Big(
	\ket{\phi^0_{12}}_A\ket{\phi^0_{12}}_B
	\bra{\phi^0_{12}}_A\bra{\phi^0_{12}}_B
	\\
	&
	+
	\ket{\Psi^+_{12}}_A\ket{\Psi^+_{12}}_B
	\bra{\Psi^+_{12}}_A\bra{\Psi^+_{12}}_B
	+
	\ket{\Psi^-_{12}}_A\ket{\Psi^-_{12}}_B
	\bra{\Psi^-_{12}}_A\bra{\Psi^-_{12}}_B
	\\
	&
	+
	\ket{\Phi^-_{12}}_A\ket{\Phi^-_{12}}_B
	\bra{\Phi^-_{12}}_A\bra{\Phi^-_{12}}_B
	\Big)
	(W_{A_1 B_1 A_2 B_2}(g,h))
	d\mu(g,h)
\end{eqnarray*}
where $\mu(\cdot,\cdot)$
is the Haar measure on $SU(2)\times SU(2)$
and where
\[
\fl	\ket{\Phi^\pm_{12}}_X
	=
	\frac
	{\ket0_{X_1}\ket0_{X_2}\pm\ket1_{X_1}\ket1_{X_2}}
	{\sqrt2},\
	\ket{\Psi^\pm_{12}}_X
	=
	\frac
	{\ket0_{X_1}\ket1_{X_2}\pm\ket1_{X_1}\ket0_{X_2}}
	{\sqrt2}
	\quad
	(X=A,B).
\]
\nashi
{
The representation $V$ is obtained by restricting $W$
to the subgroup of elements $(g,g)\in SU(2)\times SU(2)$.
Hence, a $W$-invariant projection is always
given as a mixture of $V$-invariant projections.
Indeed,
\[
	K_5^+ + K^+ + K_3^-,\
	L_3^+,\
	(\ket{\phi^0_{AB}}\bra{\phi^0_{AB}})^{\otimes2},\
	L_3^-
\]
are $W$-invariant.
Moreover,
the AB-transposition mixes $L_3^+$ and $L_3^-$.
}
By Remark \ref{rem:so3}.
a W-invariant test $T_0$ is of the form
\[
	T_0=
	w_1 (K_5^+ + K_3^- + K_1^+)
	+
	w_2 (L_3^+ + L_3^-)
	+
	w_3 \ket{\phi^0_{AB}}\bra{\phi^0_{AB}}^{\otimes2}
	.
\]
By the level-zero condition, $w_3=1$.
If $\sigma\in\cl S'$ then
\begin{equation}
	\label{eq:th:w:<}
	9^{-1}
	\tr(\sigma^{\otimes2}
	(K_5^+ + K_1^+ + K_3^-))
	\le
	6^{-1}
	\tr(\sigma^{\otimes2}
	(L_3^+ + L_3^-))
\end{equation}
because
\begin{eqnarray*}
	&&
	6^{-1}
	\tr(\sigma^{\otimes2}
	(L_3^+ + L_3^-))
	-
	9^{-1}
	\tr(\sigma^{\otimes2}
	(K_5^+ + K_1^+ + K_3^-))
	\\
	&=&
	3^{-1}
	\Big(
	x_{00}-\frac{x_{11}+x_{22}+x_{33}}{3}
	\Big)
	(x_{11}+x_{22}+x_{33})
	.
\end{eqnarray*}
As Theorem \ref{th:opt},
the type 2 error probability is uniformly minimized if
$3w_1+2w_2$ and $w_2$ are simultaneously minimized
(see Lemma \ref{lem:w-mv}).
From (\ref{eq:th1:trace}), we have
\[
	\min\{9w_1+6w_2\mid
	\mbox{AB-locality}\} = 3.
\]
Therefore $w_1=1/3$ and $w_2=w_4=0$
are the solutions to the minimization problem,
and the theorem is derived.
\hfill$\Box$

\section{Discretization of measurements}\label{sec:dis}

We have expressed
$T^u$ and $T^V$ as probabilistic mixtures
of continuously many separable operators labeled by
$SU(2)$-elements.
Such continuous expressions are simple and convenient
in a theoretical argument.
A basic method to realize a SU-invariant measurement is
to operate the system as for unitary element randomly chosen
with respect to the Haar measure.
However,
in this method,
we need to prepare continuously many operations.
Therefore, it is worth noting that
$T^u$ and $T^V$ are also expressed
as mixtures of a few operators locally realized.

\subsection{Discretization of \boldmath$T^u$}

%Let us consider the implementation of $T^u$.
%experimental situation
%
We rewrite $T$ as
\begin{eqnarray*}
	T^u_0
	&=&
	\frac23(\ket{0_A0_B}\bra{0_A0_B}+\ket{1_A1_B}\bra{1_A1_B})
	\\
	&&
	+\frac13
	(\ket{0_A1_B}\bra{0_A1_B}+\ket{1_A0_B}\bra{1_A0_B}
	+\ket{0_A1_B}\bra{1_A0_B}+\ket{1_A0_B}\bra{0_A1_B})
	\\
	&=&
	3^{-1}
	\big(
	 \ket{0_A0_B}\bra{0_A0_B}+\ket{1_A1_B}\bra{1_A1_B}
	+\ket{D_A D_B}\bra{D_A D_B}
	\\
	&&
	+\ket{X_A X_B}\bra{X_A X_B}
	+\ket{R_A L_B}\bra{R_A L_B}+\ket{L_A R_B}\bra{L_A R_B}
	\big)
\end{eqnarray*}
where
\[
	\ket{D}=\frac{\ket0+\ket1}{\sqrt2},\
	\ket{X}=\frac{\ket0-\ket1}{\sqrt2},\
	\ket{R}=\frac{\ket0+\sqrt{-1}\ket1}{\sqrt2},\
	\ket{L}=\frac{\ket0-\sqrt{-1}\ket1}{\sqrt2}
	.
\]
This means that one can realize $T^u$ by the two-values POVM
$T=\{T_0,T_1\}$
given in the form
\[
	T_0=\ket{x_A x_B}\bra{x_A x_B}+\ket{y_A y_B}\bra{y_A y_B}
\]
where the orthonormal pair $(x,y)$ is chosen from
$\{(0,1),(D,X),(R,L)\}$ completely at random.

We also note that a finite subgroup $\Bbb O$ of $SU(2)$
generated by
\[
	\left(\begin{array}{cc}\sqrt{-1}&0\\0&-\sqrt{-1}\end{array}\right)
	\mbox{ and }
	\frac{1}{\sqrt2}\left(\begin{array}{cc}1&-1\\1&1\end{array}\right)
\]
transitively acts on
$\{\ket{x_A x_B}\mid x=0,1,D,X,R,L\}$
by $U_{A B}(\cdot)$.
$\Bbb O$ is the octahedral group,
which is the (special) symmetry group of
the octahedron and the cube.
Therefore, one can also realize $T^u$ by
\[
	T_0=\ket{0_A0_B}\bra{0_A0_B}+\ket{1_A1_B}\bra{1_A1_B}
\]
after a transformation $U_{A B}(g)$ for randomly
selected $g\in\Bbb O$.

\begin{remark}
\rm
D'Ariano \etal \cite{d'ariano} have
proposed a discretization of an entanglement witness:
the same measurement as $T^u$.
Their discretized measurement is also equivalent to ours.
However,
their analysis is not enough in the sense of
hypothesis testing.
\end{remark}

\subsection{Discretization of \boldmath$T^V$}\label{sec:v-o}

%In (\ref{eq:tv:twirl}),
%$T_0^V$ is constructed as
%a classical mixture of continuously
%many local measurements.
%It means that we need to prepare
%continuously many bases to measure the system,
%so it costs much in experiments.
%To reduce the cost,
%we would like to realize
%$T_0^V$ as a mixture of finite elements
%if possible.
%In fact, it is possible.
%
%
The test $T^V$ is also expressed as a mixture of
finite measurements as follows:
\begin{equation}
	\label{eq:finitetest}
	T_0^V
	=
	\frac1{24}
	\sum_{g\in \Bbb O}
	(V_{A_1 B_1 A_2 B_2}(h^* g))^\dagger
	\Big(
	\sum_{0\le i,j\le1}
	\frac
	{ \Pi_{ij} + \tau_{12}( \Pi_{ij} ) }
	{2}
	\Big)
	V_{A_1 B_1 A_2 B_2}(h^* g)
\end{equation}
where $h^*\in SU(2)$ is defined by
\[
	h^*:
	\cos\Big(
	\frac{\arccos\sqrt{3/5}}4\Big)\ket0
	+
	\sin\Big(
	\frac{\arccos\sqrt{3/5}}4\Big)\ket1
	\mapsto\ket0
\]
and where
$\tau_{12}(\cdot)$ is
the transposition of
$A_1\otimes B_1$ and
$A_2\otimes B_2$.
%and where
%$\Bbb O$ is the finite subgroup of 24-elements generated by
%\[
%	g_1:\ket0\mapsto
%	\frac{\ket0+\ket1}{\sqrt2},\
%	g_2:\ket0\mapsto
%	\frac{\ket0+\sqrt{-1}\ket1}{\sqrt2}
%	.
%\]
%
Therefore,
one can realize $T^V$ as follows.
First, transform by $V_{A_1 B_1 A_2 B_2}(h^* g)$
where $g\in\Bbb O$ is chosen completely at random.
Next,
by probability $1/2$,
replace the sample numbering, that is, apply $\tau_{12}$.
Next,
measure the subsystems by
\[
	\begin{array}{c|c|c|c}
	A_1 & B_1 & A_2 & B_2 \\\hline
	\{\ket0,\ket1\} & \{\ket0,\ket1\} &
	\{\ket D,\ket X\} & \{\ket D,\ket X\}
	\end{array}
\]
independently.
The hypothesis $H_0$ is accepted if
$A_1$ and $B_1$ have the same measurement result
and
$A_2$ and $B_2$ have the same one.

One can check (\ref{eq:finitetest}) as follows.
The subspaces
$K_3^\pm$, $K_2^+$,
$K_1^+$, $L_1^+$ and
$L_3^\pm$
are
irreducible by the V-restriction
$\Bbb O\subset SU(2)$,
and, in particular,
the three-dimensional actions of
$\Bbb O$ for
$K_3^+$ and $L_3^+$ are
mutually inequivalent.
and, by calculation,
\begin{eqnarray}
	\label{eq:oct:1}
	\tr(K_3^+
	V_{A_1 B_1 A_2 B_2}(h_x)^\dagger
	\Pi_{ij}
	V_{A_1 B_1 A_2 B_2}(h_x))
	&=&
	\frac{\cos^2(4x)}{8},
	\\
	\label{eq:oct:2}
	\tr(K_2^+
	V_{A_1 B_1 A_2 B_2}(h_x)^\dagger
	\Pi_{ij}
	V_{A_1 B_1 A_2 B_2}(h_x))
	&=&
	\frac{\sin^2(4x)}{8}
\end{eqnarray}
where
\[
	SU(2)\ni
	h_x:
	\cos x\ket0+\sin x\ket1
	\mapsto \ket0
	,
\]
and hence
$(\ref{eq:oct:1})=(\ref{eq:oct:2})$
if $x=(\arccos\sqrt{3/5})/4$.

%Therefore,
%one can realize $T^V$ by a test $T_0=\sum_{i,j}\Pi_{i j}$
%after the $\Bbb O$-twirling.

\section{Discussion and conclusion}
\begin{figure}
\begin{center}
\includegraphics{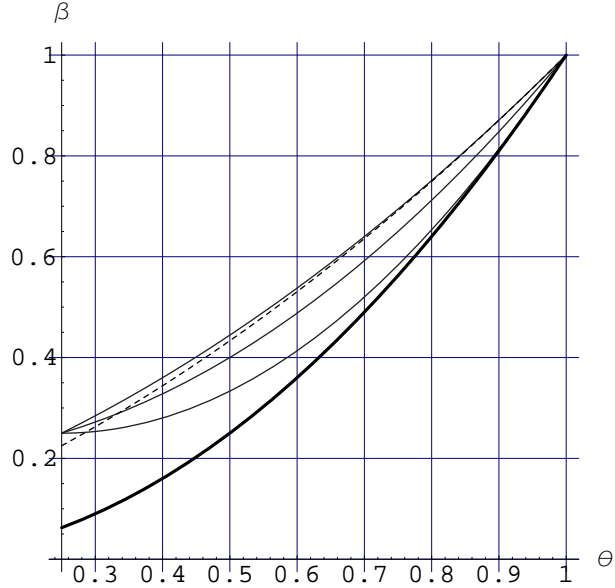}
\caption{The type 2 error probabilities $\beta$
with respect to $\theta=x_{00}=\bra{\phi^0_{AB}}\sigma\ket{\phi^0_{AB}}$
of
$T^{u,2}$ (the highest solid line),
$T^U$ (the second highest solid line),
$T^W$ (the third highest solid line),
$T^G$ (the thick line) and
$T^V$ (the dashed line)
where $x_{i j}$ are the same for $1\le i,j\le3$.}
\label{9-7}
\end{center}
\end{figure}

For $d=n=2$,
we have proposed five measurements
$T^G$, $T^{u,2}$, $T^U$, $T^V$
and $T^W$
as optimal tests
(for subsets of states, if necessary,)
in the corresponding classes of tests,
that is,
\smallskip\\
$\cl T^G$:
the class of level-zero tests,
\\
$\cl T^{u,2}$:
the class of level-zero tests
of the form $T_0^{\otimes2}$
where $T_0$ is AB-local U-invariant
for each sample,
\\
$\cl T^U$:
the class of AB-local U-invariant
level-zero tests,
\\
$\cl T^V$:
the class of AB-local, samplewise-local,
V-invariant, weakly AB-invariant,
termwise AB-covariant and level-zero tests,
%or of
%AB-local, samplewise-local, V-invariant,
%weakly AB-invariant and level-zero tests,
\\
$\cl T^W$:
the class of AB-local W-invariant
tests.
\smallskip\\
The inclusion\rev{8} relations of
these classes is not totally ordered.
For example,
from the locality,
\[
	\cl T^{u,2},\ \cl T^V
	\ \subset \
	\cl T^U, \cl T^W
	\ \subset \
	\cl T^G
\]
while from the unitary invariance,
\[
	\cl T^U
	\ \subset \
	\cl T^{u,2},\ \cl T^W
	\ \subset \
	\cl T^V
	\ \subset \
	\cl T^G
	.
\]
On the other hand,
the type 2 error probabilities of
the optimal tests are
totally ordered:
\[
	\beta(\sigma^{\otimes2},T^G)
	<
	\beta(\sigma^{\otimes2},T^W)
	<
	\beta(\sigma^{\otimes2},T^U)
	<
	\beta(\sigma^{\otimes2},T^V)
	<
	\beta(\sigma^{\otimes2},T^{u,2})
\]
in a set of states close to
$\ket{\phi^0_{AB}}$.
In Figure \ref{9-7},
the type 2 error probabilities $\beta$ are plotted
with respect to $\theta=x_{00}=\bra{\phi^0_{AB}}\sigma\ket{\phi^0_{AB}}$
of
$T^{u,2}$ (the highest solid line),
$T^U$ (the second highest solid line),
$T^W$ (the third highest solid line),
$T^G$ (the thick line) and
$T^V$ (the dashed line)
where $x_{i j}$ are the same for $1\le i,j\le3$.
If $x_{i j}=0$ for $i\ne j$,
then the line of $T^V$ coincides with that of $T^U$
and there is no change for other tests.
In such a way,
the framework of hypothesis testing
clarifies the hierarchy of
requirements for measurements
from the viewpoint of performance of
optimal tests.

We have considered hypothesis testing
for entanglement
under locality and invariance
conditions.
We have derived optimal tests
for some settings.
In our derivations of UMP tests,
the separability of LOCC measurements
played an important role.
The UMP U-invariant and level-zero test
$T^U$
was shown to have the asymptotically
same performance as $T^G$.
The PPT approach of
Virmani and Plenio \cite{virmani-plenio}
was also useful to obtain UMP tests.

We may have some problems remained.
One problem is how we can develop our results for general
level $\alpha$ ($0<\alpha<1$),
sample size $n$ and dimension $d$.
Another is what test is an appropriate test for
\begin{eqnarray}
	\nonumber
	&&
	H_0:\theta\ge c_0\mbox{ versus }H_1:\theta< c_0
	,
	\\
	&&
	H_0:\theta\le c_0\mbox{ versus }H_1:\theta> c_0
	\label{eq:sighypo}
\end{eqnarray}
for a constant $c_0$ very close to one.
Indeed, if $H_0$ of (\ref{eq:sighypo}) 
is rejected by a test with small level,
then the statement `The state is very close to $\ket{\phi^0_{AB}}$'
will be strongly supported.
Hence, it is siginificant to treat 
the hypothesis of the form (\ref{eq:sighypo}).
This problem will be treated in a forcoming paper \cite{newhayashi}.
It is also a problem remained
to remove technical assumptions
such as (\ref{eq:ineqwithout}) in Section 5.2.

\ack
The authors thank the referees for useful comments.

\appendix
\section{Lemmas for Theorems \ref{th:opt}, \ref{th:w}}

\begin{lemma}\label{lem:5w1+3w2}
(For Theorem \ref{th:opt}) \
If $1-\theta$ is small enough,
then
the power of the test in Theorem \ref{th:opt} is
uniformly maximized if (\ref{eq:5w1+3w2})
is simultaneously maximized.
\end{lemma}

\noindent
{\bf Proof} \ \
Since $K_i^\pm$ is spanned by $x_{i j}$ and
$L_i^\pm$ is spanned by $x_{i,0}$ and $x_{0,j}$
for $1\le i,j\le 3$,
it holds that
\begin{equation}
	\tr(\sigma^{\otimes2}K_i^+)=O\big((1-\theta)^2\big)
	\ \mbox{ and } \
	\tr(\sigma^{\otimes2}L_i^+)=O(1-\theta)
	\label{eq:proj-order}
\end{equation}
as $\theta\to1-0$,
except for $L_1^+$.
By (\ref{eq:tr:K_5^+}) and (\ref{eq:tr:L_3^+}),
\begin{eqnarray}
	\nonumber
	\tr\big(\sigma^{\otimes2}(3K_5^+-5K_3^-)\big)
	&=&
	(x_{11}-x_{22})^2+(x_{22}-x_{33})^2+(x_{33}-x_{11})^2
	\\
	&&
	+4({\rm Im}\, x_{12})^2
	+4({\rm Im}\, x_{23})^2
	+4({\rm Im}\, x_{31})^2
	\nonumber
	\\
	&&
	+6(|x_{12}|^2+|x_{23}|^2+|x_{31}|^2)
	\nonumber
	\\
	&\ge&0
	.
	\label{eq:s5>l3+}
\end{eqnarray}
%
%\begin{equation}
%	\label{eq:s5>l3+}
%	\tr\big(\sigma^{\otimes2}(K_5^+-K_3^-)\big)
%	=
%	2 \sum_{i=1}^3 x_{i,i}^2
%	+3\sum_{i\ne j}|x_{i,j}|^2
%	-\sum_{i\ne j}x_{i,j}^2
%	\ge0.
%\end{equation}
Define column vectors $v$ and $w$ by
\begin{eqnarray*}
	&&
	v=
	\left(\begin{array}{ccccc}
	\tr(\sigma^{\otimes2}K_5^+) &
	\tr(\sigma^{\otimes2}L_3^+) &
	\tr(\sigma^{\otimes2}K_1^+) &
	\tr(\sigma^{\otimes2}K_3^-) &
	\tr(\sigma^{\otimes2}L_3^-) 
	\end{array}\right)^T
	,
	\\
	&&
	w=
	\left(\begin{array}{ccccc}
	w_1 & w_2 & w_3 & w_4 & w_5
	\end{array}\right)^T
	,
\end{eqnarray*}
and define a $5\times5$ matrix $M$ by
\begin{eqnarray*}
\fl	\quad
M=
	\frac{1}{15}
	\left(\begin{array}{ccccc}
	3 & -9 & 0 & 0 & 0 \cr
	0 & 15 & 0 & 0 & 0 \cr
	0 & 0 & 15 & 0 & 0 \cr
	-5 & 0 & -5 & 5 & -15 \cr
	0 & 0 & 0 & 0 & 15
	\end{array}\right)
	\quad\mbox{ as the inverse of }\quad
	\left(\begin{array}{ccccc}
	5 & 3 & 0 & 0 & 0 \cr
	0 & 1 & 0 & 0 & 0 \cr
	0 & 0 & 1 & 0 & 0 \cr
	5 & 3 & 1 & 3 & 3 \cr
	0 & 0 & 0 & 0 & 1
	\end{array}\right)
	.
\end{eqnarray*}
Let $v'=M^T \cdot v$ and $w'=M^{-1}\cdot w$.
Each quantity in (\ref{eq:5w1+3w2}) is an entry of $w'$,
and
the power of the test in Theorem \ref{th:opt} is
given as $v^T \cdot w={v'}^T\cdot w'$.
When each entry of $v'$ is non-negative,
the maximum of ${v'}^T \cdot w'$ is attained by
maximizing $w'$.
From (\ref{eq:s1>l3+}), (\ref{eq:proj-order}),
and (\ref{eq:s5>l3+}),
there is $\theta_0$ such that
$v'$ is non-negative.
Therefore,
if (\ref{eq:5w1+3w2}) is maximized,
then the power of the test is maximized.
\hfill$\Box$

\medskip

Let ${\rm pt}_C(X)$ be the partial transpose of
an operator $X$ on a subsystem $C$, for example,
${\rm pt}_{A_1\otimes B_1}(X)$ is given by
\begin{eqnarray*}
\fl \quad	{\rm pt}_{A_2 B_2}(X)=
	\sum_{0\le i,j,k,l \le1}
	I_{A_1 B_1} \otimes
	\ket{i}_{A_2}\ket{j}_{B_2}
	\bra{k}_{A_2}\bra{l}_{B_2}
	X
	I_{A_1 B_1} \otimes
	\ket{i}_{A_2}\ket{j}_{B_2}
	\bra{k}_{A_2}\bra{l}_{B_2}
\end{eqnarray*}
where $I_{A_2 B_2}$ is the identity on $A_2\otimes B_2$.

\begin{lemma}\label{lem:1}
(For Theorem \ref{th:opt}) \
If $T=\{T_0,T_1\}$ is samplewise-local then
$w_2=w_5$.
\end{lemma}
{\bf Proof} \ \
The samplewise-locality of $T$ implies that
${\rm pt}_{A_2\otimes B_2}(T_1)$ is positive,
in particular,
\begin{eqnarray*}
\fl	R=
	\left(\begin{array}{cc}
	\bra{u} {\rm pt}_{A_2\otimes B_2}(T_1) \ket{u} &
	\bra{u} {\rm pt}_{A_2\otimes B_2}(T_1) \ket{v} \cr
	\bra{v} {\rm pt}_{A_2\otimes B_2}(T_1) \ket{u} &
	\bra{v} {\rm pt}_{A_2\otimes B_2}(T_1) \ket{v}
	\end{array}\right)
	=
	\left(\begin{array}{cc}
	0 & -\frac{5\sqrt{-1}}{6\sqrt3} (w_2-w_5) \cr
	\frac{5\sqrt{-1}}{6\sqrt3} (w_2-w_5) &
	\frac{17 w_1 + 9 w_3 + w_4}{27}
	\end{array}\right)
\end{eqnarray*}
should be positive where
\begin{eqnarray*}
	\ket{u}=
	\ket{\phi^0_{AB}}_1 \ket{\phi^0_{AB}}_2,\
	\ket{v}=
	\frac
	{
		5
		\ket{\phi^1_{AB}}_1\ket{\phi^1_{AB}}_2
		-
		\ket{\phi^2_{AB}}_1\ket{\phi^2_{AB}}_2
		-
		\ket{\phi^3_{AB}}_1\ket{\phi^3_{AB}}_2
	}
	{3\sqrt3}
	.
\end{eqnarray*}
Since $\det(R)=-25/108 (w_2-w_5)^2\ge0$ holds,
$w_2=w_5$.
\hfill$\Box$

\begin{lemma}\label{lem:2}
(For Theorem \ref{th:opt}) \
If $T=\{T_0,T_1\}$ is AB-local then
\begin{eqnarray}
	0 \le \frac{10w_1+6w_2-w_3}{12} \le 1,
	\label{eq:lem:1}	\\
	0 \le \frac{w_3+2(w_4+w_5)}{4} \le 1
	.	\label{eq:lem:2}
\end{eqnarray}
\end{lemma}
{\bf proof}\ \
The AB-locality of $T$ implies that
${\rm pt}_{B_1\otimes B_2}(T_0)=
{\rm pt}_{B_1\otimes B_2}(I-T_1)$ is positive.
The first result (\ref{eq:lem:1}) is obtained since
\begin{eqnarray*}
\fl \quad	&&
	\frac12
	(
	\bra{0_{A_1} 1_{B_1} 0_{A_2} 1_{B_2}} -
	\bra{1_{A_1} 0_{B_1} 1_{A_2} 0_{B_2}} )
	{\rm pt}_{B_1\otimes B_2}(T_1)
	(
	\ket{0_{A_1} 1_{B_1} 0_{A_2} 1_{B_2}} -
	\ket{1_{A_1} 0_{B_1} 1_{A_2} 0_{B_2}} )
	\\
\fl	&
	=&
	\frac{10w_1+6w_2-w_3}{12}
	.
\end{eqnarray*}
The second result (\ref{eq:lem:2}) is obtained since
\begin{eqnarray*}
\fl \quad	&&
	\frac12
	(
	\bra{0_{A_1} 0_{B_1} 0_{A_2} 0_{B_2}} -
	\bra{1_{A_1} 1_{B_1} 1_{A_2} 1_{B_2}} )
	{\rm pt}_{B_1\otimes B_2}(T_1)
	(
	\ket{0_{A_1} 0_{B_1} 0_{A_2} 0_{B_2}} -
	\ket{1_{A_1} 1_{B_1} 1_{A_2} 1_{B_2}} )
	\\
\fl	&
	=&
	\frac{w_3+2(w_4+w_5)}{4}
	.
\end{eqnarray*}
\hfill$\Box$

\begin{lemma}\label{lem:3}
(For Theorem \ref{th:opt}) \
If $T=\{T_0,T_1\}$ is AB-local and samplewise-local then
\begin{eqnarray*}
	\frac{3}{4}(w_2+w_5) \le 1.
\end{eqnarray*}
\end{lemma}
{\bf Proof} \ \
The AB-locality and samplewise-locality of $T$ implies that
${\rm pt}_{B_2}(T_0)=
{\rm pt}_{B_2}(I-T_1)$ is positive.
Since
\begin{eqnarray*}
	\bra{\phi^0_{AB}}_1\bra{\phi^2_{AB}}_2
	{\rm pt}_{B_2}(T_1)
	\ket{\phi^0_{AB}}_1\ket{\phi^2_{AB}}_2
	=
	\frac34(w_2+w_5),
\end{eqnarray*}
we have the result.
\hfill$\Box$

\begin{lemma}\label{lem:w-mv}
(For Theorem \ref{th:w}) \
In Theorem \ref{th:w},
the type 2 error probability of the test is
uniformly minimized
if $3w_1+2w_2$ and $w_2$ are simultaneously minimized.
\end{lemma}
{\bf Proof} \ \
Define column vectors $v$ and $w$ by
\begin{eqnarray*}
	&&
	v=
	\left(\begin{array}{cc}
	\tr(\sigma^{\otimes2}(K_5^++K_1^++K_3^-)) &
	\tr(\sigma^{\otimes2}(L_3^++L_3^-))
	\end{array}\right)^T
	,
	\\
	&&
	w=
	\left(\begin{array}{cc}
	w_1 & w_2
	\end{array}\right)^T
	,
\end{eqnarray*}
and define a $2\times2$ matrix $M$ by
\begin{eqnarray*}
	M=
	\frac{1}{3}
	\left(\begin{array}{cc}
	1 & -2 \cr
	0 & 3
	\end{array}\right)
	\quad\mbox{ as the inverse of }\quad
	\left(\begin{array}{cc}
	3 & 2 \cr
	0 & 1
	\end{array}\right)
	.
\end{eqnarray*}
Let $v'=M^T\cdot v$ and $w'=M^{-1}\cdot w$.
The power of the test in Theorem \ref{th:w} is
given as $v^T\cdot w={v'}^T\cdot w'$.
If each entry of $v'$ is non-negative,
the maximum of ${v'}^T\cdot w'$ is attained by
maximizing $w'$.
From (\ref{eq:th:w:<}),
$v'$ is non-negative.
Therefore,
if $3w_1+2w_2$ and $w_2$ are simultaneously minimized,
the type 2 error probability is minimized.
\hfill$\Box$

\section*{References}
%\References


\begin{thebibliography}{99}

%\bibitem{acin-tarrach-vidal}
%Acin, A., Tarrach, R. and Vidal G.
%2000
% Optimal estimation of two-qubit pure-state entanglement
%{\it Phys. Rev. A} {\bf 61}, 62307.

\bibitem{barbierietal}
Barbieri, M., De Martini, F., Di Nepi, G., Mataloni, P.,
D'Ariano, G. M., and Macchiavello, C.
2003
%Detection of Entanglement with Polarized Photons:
%Experimental realization of an Entanglement Witness,
{\it Phys. Rev. Lett.} {\bf 91} 227901.


\bibitem{bbcjpw}
Bennett, C., Brassard, G., Crepeau, C.,
Jozsa, R., Peres, A., and Wootters,  W.K.
1993
{\it Phys. Rev. Lett.} {\bf 70} 1895.

\bibitem{bennett-wiesher}
Bennett, C. and Wiesner, S.J. 
1992
{\it Phys. Rev. Lett.} {\bf 69} 2881.

\bibitem{bhm}
Brown, L. D., Hwang, J. T. G. and Munk, A.
1997
%An unbiased test for the bioequivalence problem.
{\it Ann. Statist.} {\bf 25} 2345-2367.


\bibitem{d'ariano}
D'Ariano, G. M., Macchiavello, C. and Paris M. G. A.
2003
%Local observables for entanglement witnesses
{\it Phys. Rev. A} {\bf 67} 042310.

\bibitem{ekert}
Ekert, A.
1991
{\it Phys. Rev. Lett.} {\bf 67} 661.

\bibitem{fulton}
Fulton, W. and Harris, J.
1991
{\it Representation Theory; A First Course},
Springer, New York.


\bibitem{goodman}
Goodman, R. and Wallach, N. R.
1998
{\it Representations and invariants of the classical groups},
Cambridge University Press, Cambridge.

\bibitem{guhne}
G\"uhne, O., Hyllus, P., Brus, D., Ekert, A., Lewenstein, M.,
Macchiavello, C. and Sanpera, A.
2002
%Detection of entanglement with few local measurements
{\it Phys. Rev. A} {\bf 66} 062305.

\bibitem{hayashi-hypo}
Hayashi, M. 2002
%Optimal sequence of quantum measurements in the sense of Stein's lemma in 
%quantum hypothesis testing. 
{\it J. Phys. A} {\bf 35} 10759-10773.

\bibitem{asympt}
Hayashi, M.
2005
{\it Asymptotic Theory of Quantum Statistical Inference:
Selected Papers}
World Scientific.

\bibitem{newhayashi}
Hayashi, M.
%2005
%Hypotheses testing for maximally entangled state
%To be submitted.
in preparation.

\bibitem{hayashi-text}
Hayashi, M.
2006
{\it Quantum Information: An Introduction},
Springer.

\bibitem{damian}
Hayashi, M, Markham, D., Murao, M., Owari, M., and  Virmani, S.
2006
{\em Phys. Rev. Lett.} {\bf 96} 040501.

\bibitem{helstrom}
Helstrom, C. W.
1976
{\it Quantum detection and estimation theory}
Academic Press.

\bibitem{hiai-petz}
Hiai, F. and Petz, D. 1991
%The proper formula for relative entropy and its asymptotics in quantum
%probability.
{\it Comm. Math. Phys.}, {\bf 143}, 99-114.


\bibitem{holevo}
 Holevo, A. S.
1982
{\it Probabilistic and statistical aspects of quantum theory},
North-Holland.

\bibitem{horodecki3}
Horodecki, M., Horodecki, P. and Horodecki, R.
1996
{\it Phys. Lett. A} {\bf 223} 1.

\bibitem{lehmann}
Lehmann, E. L.
1986
{\it Testing statistical hypotheses},
Second edition.
Wiley.


\bibitem{lewenstein}
Lewenstein, M., Kraus, B., Cirac, J.I. and Horodecki, P. 
2000
{\it Phys. Rev. A} {\bf 62} 052310.

\bibitem{nielsen-chuang}
Nielsen, M. A. and Chuang, I. L.
2000
{\it Quantum computation and quantum information}
Cambridge University Press.


\bibitem{rains1}
Rains, E.M.
1999
{\it Phys. Rev. A} {\bf 60} 173.

\bibitem{rains2}
Rains, E.M.
2001
{\it IEEE Trans. Inf. Theory} {\bf 47} 2921.

%\bibitem{shannon}
%Shannon, C. E.
%1948
%A mathematical theory of communication.
%{\it Bell System Tech. J.} {\bf 27} 379.


\bibitem{terhal}
Terhal, B. M.
2000
%Bell Inequalities and the Separability Criterion
{\it Phys. Lett. A} {\bf 271} 319.


\bibitem{ogawa-nagaoka}
Ogawa, T. and Nagaoka, H. 2000
%Strong converse and Stein's lemma in quantum hypothesis testing. 
{\it IEEE Trans. Inform. Theory} {\bf 46} 2428-2433.

\bibitem{owari}
Owari, M. and Hayashi, M
2005
quant-ph/0509062; to appear in {\it Phys. Rev. A}.

\bibitem{virmani-plenio}
Virmani, S. and Plenio, M. B.
2003
%Construction of extremal local positive-operator-valued measures under 
%symmetry.
{\it Phys. Rev. A} {\bf 67} 062308.

\end{thebibliography}
\end{document}